\documentclass[a4paper,fleqn,usenatbib]{mnras}

\usepackage[T1]{fontenc}
\usepackage{ae,aecompl}

\usepackage{amsmath}
\usepackage{amssymb}
\usepackage{graphicx}
\usepackage{multirow}
\usepackage[hang, flushmargin]{footmisc}
\usepackage{bm}
\usepackage{hyperref}
\usepackage{color}
\usepackage{textcomp}
\usepackage{url}
\usepackage[normalem]{ulem}
\usepackage{float}
\usepackage{fancyvrb}
\usepackage{pgfplotstable}
\usepackage{booktabs}
\usepackage{multicol}
\usepackage{pdflscape}
\usepackage{courier}
\usepackage{txfonts}

\DefineVerbatimEnvironment{code}{Verbatim}{fontsize=\small}
\DefineVerbatimEnvironment{example}{Verbatim}{fontsize=\small}
\renewcommand{\micron}{\hbox{\textmu}m}
\newcommand{\ujy}{\hbox{\textmu}Jy}

\title[M31 observed with \emph{Spitzer}-IRAC]{The extended disc and halo of the Andromeda galaxy observed with \emph{Spitzer}-IRAC}

\author[M.~Rafiei~Ravandi et al.]{
Masoud~Rafiei~Ravandi,$^{1}$\thanks{mrafieir@uwo.ca (MRR)}
Pauline~Barmby,$^{1,2}$\thanks{pbarmby@uwo.ca (PB)}
Matthew~L.\,N.\ Ashby,$^{3}$
Seppo~Laine,$^{4}$
\newauthor
T.\,J.~Davidge,$^{5}$
Jenna~Zhang,$^{3}$
Luciana~Bianchi,$^{6}$
Arif~Babul,$^{7}$
S.\,C.~Chapman$^{8}$
\\
$^{1}$Department of Physics and Astronomy, Western University, 1151 Richmond Street, London, ON, N6A~3K7, Canada\\
$^{2}$Centre for Planetary Science and Exploration, Western University, 1151 Richmond Street, London, ON, N6A~3K7, Canada\\
$^{3}$Harvard-Smithsonian Center for Astrophysics, 60 Garden St., Cambridge, MA~02138, USA\\
$^{4}$Spitzer Science Center, MS 314-6, California Institute of Technology, 1200 East California Blvd, Pasadena, CA~91125, USA\\
$^{5}$Dominion Astrophysical Observatory, National Research Council of Canada, 5071 West Saanich Road, Victoria, BC, V9E~2E7, Canada\\
$^{6}$Department of Physics \& Astronomy, The Johns Hopkins University, 3701 San Martin Drive, Baltimore, Maryland~21218, USA\\
$^{7}$Department of Physics and Astronomy, Elliott Building, University of Victoria, 3800 Finnerty Road, Victoria, BC, V8P~5C2, Canada\\
$^{8}$Department of Physics and Atmospheric Science, Dalhousie University, Halifax, NS, B3H~4R2, Canada
}

\date{Accepted 2016 March 14. Received 2016 March 14; in original form 2015 December 15}

\pubyear{2016}
\graphicspath{ {./Figs/} }
\begin{document}
\label{firstpage}
\pagerange{\pageref{firstpage}--\pageref{lastpage}}
\maketitle

\begin{abstract}
We present the first results from an extended survey of the Andromeda galaxy (M31) using 41.1~h of
observations by {\it{Spitzer}}-IRAC at 3.6 and 4.5\,\micron. This survey extends previous observations
to the outer disc and halo, covering total lengths of 4\fdg4 and 6\fdg6 along the minor and major axes,
respectively. We have produced surface brightness profiles by combining the integrated light from background-corrected
maps with stellar counts from a new catalogue of point sources. Using auxiliary catalogues we have carried out a
statistical analysis in colour-magnitude space to discriminate M31 objects from foreground Milky Way stars and
background galaxies. The catalogue includes 426,529 sources, of which 66~per~cent have been assigned probability values to identify M31 objects with magnitude depths of [3.6]$\,=\,$19.0$\,\pm\,$0.2, [4.5]$\,=\,$18.7$\,\pm\,$0.2. We discuss applications of our data for constraining the stellar mass and characterizing point sources in the outer radii.
\end{abstract}


\section{Introduction}
Located at a distance of mere $\sim\,$800\,kpc, the Andromeda galaxy (M31) is our
closest laboratory for carrying out a `galaxy dissection' in the Local Group. 
M31 ($0^{\rm h}42^{\rm m}44^{\rm s}, +41\degr16\arcmin8\arcsec$) is the large galaxy for which we have the best hope of understanding all
components from an external perspective, providing an important and unique comparison
to models and observations of high-redshift galaxies. Recent optical surveys \citep{guha05, ibata07, ibata14}
have shown that the disc and halo of M31 extend much further than previously thought,
and contain stellar streams and arcs which reveal the complex history of this galaxy.
The {\it{GALEX}} discovery of star-formation in outer discs \citep{Thilker07, bianchi09},
show that outer galaxy discs and halos contain important clues about disc formation, star
formation, halo formation and galaxy evolution at all epochs. These clues were
not accessible with previous studies of `classical' galaxy properties within the `optical diameter', $D_{25}$.

The most recent studies \citep{geehan06, courteau11} suggest a full three-component picture of M31,
consisting of a bulge, a disc, and an extended halo. Probing the bright, crowded stellar bulge is
best done in integrated light while the distant halo 
can be detected only with star counts. The stellar disc is probed by both the integrated surface 
brightness profile  and star counts. The old, low-mass stars which make up the bulk of the stellar mass
\citep{rix93} have their spectral energy distribution (SED) peaks in the near-infrared band,
where the effects of extinction are greatly reduced compared to visible light.
This is especially important in a highly-inclined galaxy such as M31
($i\,$=\,77${^\circ}$, \citealp{corbelli}).

Though M31 is one of the most extensively studied galaxies, many of its structural parameters remain uncertain.
Near-infrared 2MASS `6x' imaging of M31 has been used to clarify the nature of M31's boxy bar and
bulge \citep{AB06,Beaton07}; however 2MASS imaging is too shallow to probe the outer regions of the galaxy.
While combining near-infrared data with visible light star counts is one way to
bridge the gap between the inner and outer disc, such combinations suffer from uncertainty about whether the
same stellar populations are being probed. Using an $I$-band composite profile, \citet{courteau11} computed a S\'{e}rsic 
index value of $n\simeq2.2\pm0.3$ for the bulge. Depending on the passbands and fitting techniques in use,
disc scale lengths can vary by $\sim\,$20~per~cent. Results for the halo parameters are more promising, with a general
agreement on a power-law index of $-2.5\pm0.2$
\citep{irwin05, ibata07, tanaka10, courteau11}. The bulge and inner disc parameters are not independent
of the outer disc and inner halo \citep{geehan06}. Therefore, an extended multiwavelength picture of the galaxy can help us
better constrain all the parameters of the three components.

Simultaneous measurements of the outer disc and halo are extremely difficult for large galaxies observed in the ground-based near-infrared. The highly-variable near-infrared sky and the time needed to map large
angular distances to a background region make background levels highly uncertain (e.g., the 2MASS Large
Galaxy Atlas image of M31 is known to suffer from background subtraction problems, \citealp{barmby06}).
The assumed background level can have huge effects on a derived surface brightness profile
\citep{fingerhut10,sheth10}: an over-subtracted background makes the profile too steep while an
under-subtracted one makes it too shallow.

Mid-infrared imaging with  {\it Spitzer}'s Infrared Array Camera \citep[IRAC,][]{fazio04} provides a new strategy to trace the stellar mass of M31 to large radii.
The efficiency of {\it Spitzer} in mapping large fields and the dark infrared background in space are
key features not available from ground-based facilities.
The mid-infrared bands trace both the integrated light of low-mass stars and bright asymptotic giant branch (AGB) stars with circumstellar envelopes, providing a unique opportunity to constrain and calibrate stellar populations in colour-magnitude space.
In this paper, we present new 3.6 and 4.5\,\micron\ observations of M31 which extend to large projected distances along the
major and minor axes.
These data are combined with earlier observations of the M31 disc and the resulting mosaics
carefully background-subtracted. The extended mosaics are used to produce a point-source catalogue
and measure the surface brightness profile of the galaxy in integrated light.
A statistical study of point sources in colour-magnitude space is used to extend the galaxy profile via star counts.
Using the combined surface brightness profile we assess the applications of our data for finding the best physical model of the galaxy.

We adopt a distance of $785\pm25$\,kpc \citep[DM$\,=\,$$24.47\pm0.07$,][]{mcdist05} to M31. An inclination angle of $i\,$=\,77${^\circ}$ and a position angle of 37\fdg715 (East of North) are assumed.
The Vega magnitude system is adopted with, e.g., [3.6] indicating an apparent magnitude at 3.6\,\micron,
and M$_{3.6}$ indicating an absolute magnitude.
M31-galactocentric distances are deprojected (i.e., along the major axis), unless stated otherwise.

\section{Observations and Data Reduction}
\subsection{Mapping strategy and calibration}
\label{sec:mapping}
\begin{figure}
\centering
\includegraphics[height=108mm]{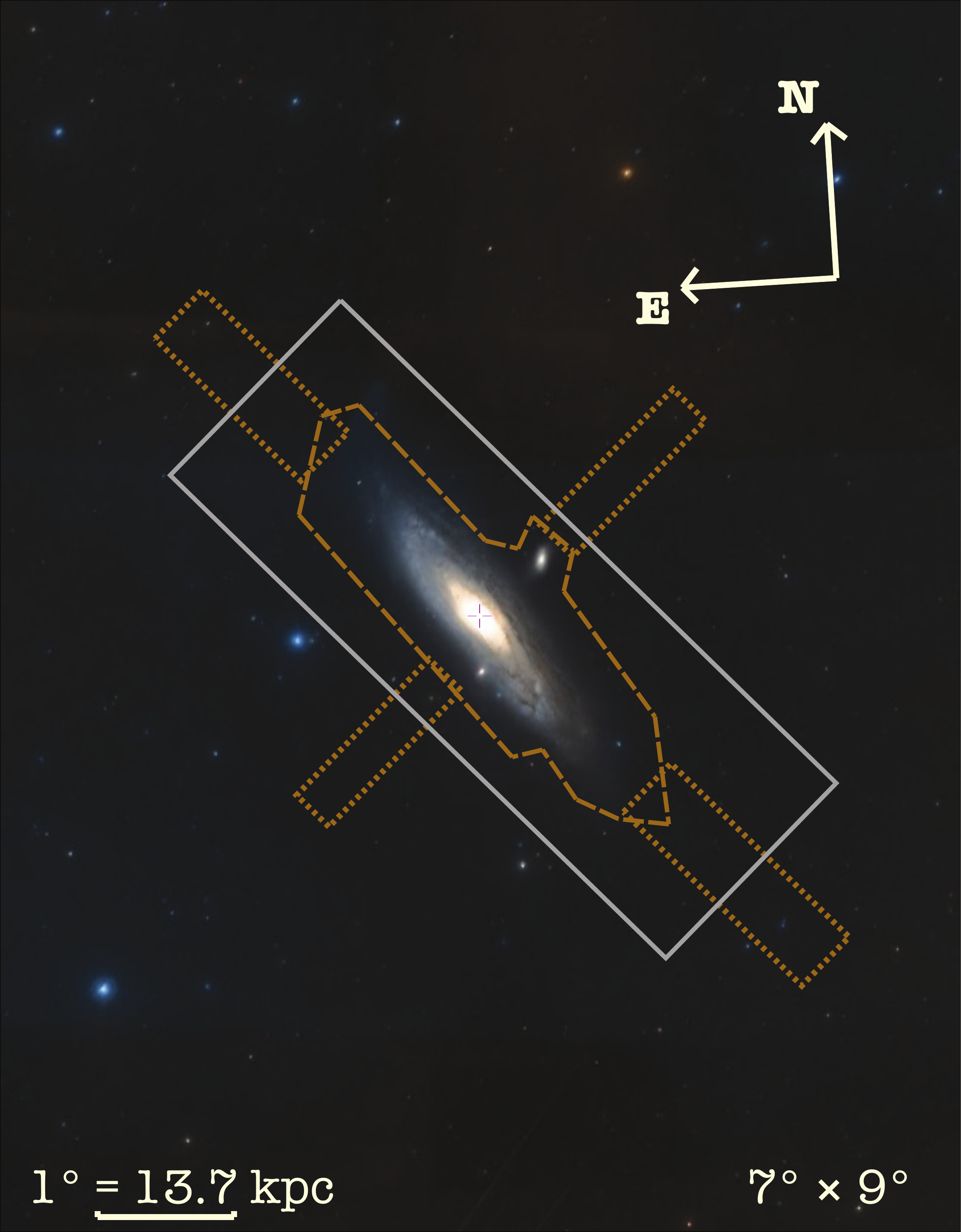}
\caption{Digitized Sky Survey coloured map showing the extent of M31 IRAC coverage (long-dashed lines, \citealp{barmby06}), 
MIPS coverage (solid lines, $5\fdg1 \times 1\fdg8$, \citealp{gordon06}), and the extended observations (dotted lines).}
\label{fig:m31regions}
\end{figure}

This work is based on 41.1\,hrs of observations at 3.6\,\micron\ and 4.5\,\micron\ by
IRAC during \emph{Spitzer}'s warm-mission Cycle 8 with program ID (80032), between September and November 2012.
Fig.~\ref{fig:m31regions} shows the extended regions that slightly overlap with data from Cycle 1 \citep{barmby06}.
The total coverage of the available data is 4\fdg4 along the minor axis and 6\fdg6 along the major axis.
The optical diameter of M31 is $D_{25}\,$$=$\,3\fdg4 \citep{rc3} so the new observations reach
about $2 \,\times D_{25}$. The observations are summarized in Table~\ref{tab:obs}.

For the observation depth, we followed the {\em Spitzer} Survey of Stellar Structures in 
Galaxies \citep[S$^4$G,][]{sheth10}  and specified eight 30-second frames per sky position. 
This exposure time is twice that of the IRAC Cycle-1 observations of M31, which is
appropriate because we were attempting to trace fainter structures.
Each sky position was covered in two astronomical observation requests (AORs), each of which contained four dithered observations per position.
The two AORs were observed a few hours or days apart to enable better rejection of asteroids.

\begin{table*}
\centering
\caption{Observation Summary}
\label{tab:obs}
\begin{tabular}{lcccccr}
\hline\hline
Target Field & Position & Area & AORID & Observation Date\\
(Axis) & (J2000) & & &\\
\hline
Major - NE & 00:50:20 +43:05:00 & $30^\prime \times 90^\prime$ & 42273024 & 2012 Oct 31\\
& & & 42274048 & 2012 Nov 03\\
Major - SW & 00:33:34 +39:20:00 & $30^\prime \times 90^\prime$ & 42273792 & 2012 Oct 23\\
& & & 42272768 & 2012 Oct 25\\
Minor - NW & 00:48:05 +40:20:21 & $20^\prime \times 90^\prime$ & 42274304 & 2012 Oct 22\\
& & & 42273280 & 2012 Oct 26\\
Minor - SE & 00:36:41 +42:47:20 & $20^\prime \times 105^\prime$ & 42273635 & 2012 Sep 22\\
 & & & 42274560 & 2012 Sep 22\\
\hline
\end{tabular}
\end{table*}

The data reduction was based on Corrected Basic Calibrated Data (CBCD) files created by
version S18.25.0 of the ground-system pipeline.  The CBCD exposures were examined individually to exclude files
compromised by scattered light and solar cosmic rays.
All nominal exposures were processed using custom routines to correct column-pulldown effects
arising from bright sources.  The code used for this purpose, known as the ``Warm-Mission Column Pulldown Corrector,"
is publicly available in the NASA/IPAC Infrared Science Archive (IRSA).\footnote{\url{http://irsa.ipac.caltech.edu/data/SPITZER/docs/dataanalysistools/tools/contributed/irac/fixpulldown/}}

After these preliminaries, all the 3.6 and 4.5\,\micron\ exposures were separately co-added into monolithic mosaics
using IRACproc \citep{IRACproc}. IRACproc implements the standard IRAC data reduction software \citep[{\sc{mopex}},][]{mopex}
with a refinement that correctly handles outlier rejection in pixels having high surface-brightness gradients;
this is done to account for the slightly under-sampled IRAC point spread function in 3.6 and 4.5\,\micron\ passbands.
The software was configured to automatically flag and reject cosmic ray artefacts. 
The output mosaics were generated with 1\farcs2 pixels, close to the native IRAC pixel scale. This scale slightly under samples
the angular resolution of the telescope and instrument at these wavelengths (1\farcs8).
Each mosaic was paired with an associated coverage map that
indicated the number of IRAC exposures used to construct each mosaic pixel.  Because of limitations imposed by computer memory,
two sets of mosaic-coverage-map pairs were generated.  One set was aligned with the galaxy major axis, covering
a narrow strip 50\arcmin$\times\,$6\fdg5 centred on M31.  The other set was aligned with the galaxy minor axis,
covering an overlapping orthogonal strip 25\arcmin$\times\,$5\degr.
Combining the IRAC exposures from the warm and cryogenic missions thus results in continuous coverage of both the major
and minor axes of this galaxy that extends several scale lengths beyond the visible galaxy disc.
Because of dithering, combination of data from multiple AORs, and the offset fields of view in the two bands, the depth of observations varies 
with location in the final mosaic.
The central region of the galaxy covered with 6$\times$12~s exposures extends for approximately 100\arcmin\,on the major axis;
beyond this, the outer part of the Cycle 1 observations, with 4$\times$30~s exposures, extends for $\sim\,$50\arcmin\,to North and South;
the area covered in the new observations, with 8$\times$30~s exposures, extends  for $\sim\,$80\,-\,90\arcmin\,again to both North and South.

\subsection{Flux uncertainty estimation}
\label{sec:bg_unc}
As part of the main pipeline, mosaics were corrected for the instrument
bias and the (much smaller) dark current.
It was assumed that all of the darks are noise-free, and the non-linear background is entirely due to
the zodiacal light \citep{zodiac12}. An absolute calibration uncertainty is computed by multiplying an estimate for
the zodiacal light, provided in the BCD header as \texttt{ZODY\char`_EST},
by the extended-to-point source calibration ratios (with 10~per~cent uncertainty) given in the IRAC Instrument Handbook.
The value of  \texttt{ZODY\char`_EST} is computed by a model with 2~per~cent uncertainty. Based on the upper limits of
\texttt{ZODY\char`_EST} in a few randomly picked BCD images over the entirety of observed fields,
we found absolute calibration uncertainty values of $\sigma_{zod}=$ 0.0014 and 0.006\,MJy~sr$^{-1}$ at 3.6 and 4.5\,\micron , respectively.
The combined uncertainty value of a given pixel is therefore $(\sigma_{p}^2+\sigma_{zod}^2)^{1/2}$\,, where $\sigma_{zod}$ is in electron counts and $\sigma_{p}$ is the Poisson noise associated with random photons.

\subsection{Background subtraction}
\label{sec:bg_sub}
\begin{figure}
\centering
\includegraphics[height=90mm]{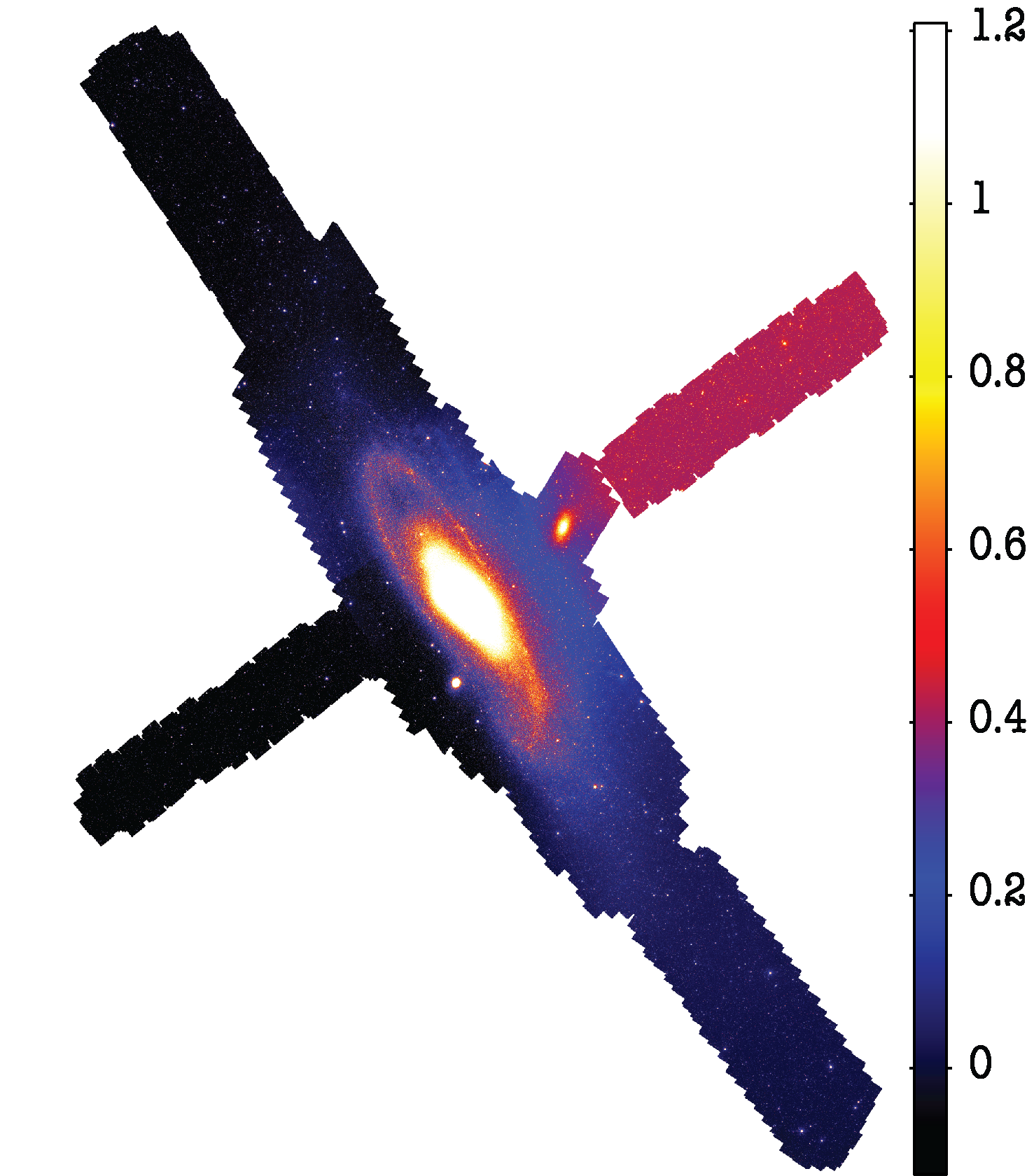}
\caption{Calibrated raw mosaic of M31, observed at 3.6\,\micron\ during the cryogenic and warm cycles of \emph{Spitzer}-IRAC.
  North points up and East is towards left.
  The colourbar [MJy~sr$^{-1}$] is constrained and scaled to show background variations.
  {\sc{montage}} \citep{montage} was used for stitching the two minor- and major-axis mosaics.}
\label{fig:raw_colour}
\end{figure}

Accurate background subtraction is critical for surface photometry at faint levels.
The `first-frame effect' in the IRAC detectors combines with the instrument's lack of a shutter to make background levels highly uncertain.
Background-matching on local scales in the mosaicing process can induce significant gradients in
background over large scales \citep{arendt10}. 
Fig.~\ref{fig:raw_colour} shows the large background gradient in the 3.6\,\micron\ 
raw mosaic; this gradient, if unsubtracted, would swamp any signal from the low-surface-brightness outer disc. 
To model the background, mosaic regions covered by M31 and its
satellites were heavily masked up to where point sources could be distinguished from one another
(i.e., uncrowded regions). Source~Extractor \citep[{\sc{SExtractor}},][]{sextractor} was used to locate point sources and other background discontinuities for further masking. {\sc{SExtractor}} parameters were fine-tuned by visual inspection so that a significant number of regions were masked -- see section~\ref{sec:irac_cat} for a full discussion on the robust detection and photometry of IRAC sources using {\sc{SExtractor}}.

\begin{figure}
\centering
\includegraphics[height=90mm]{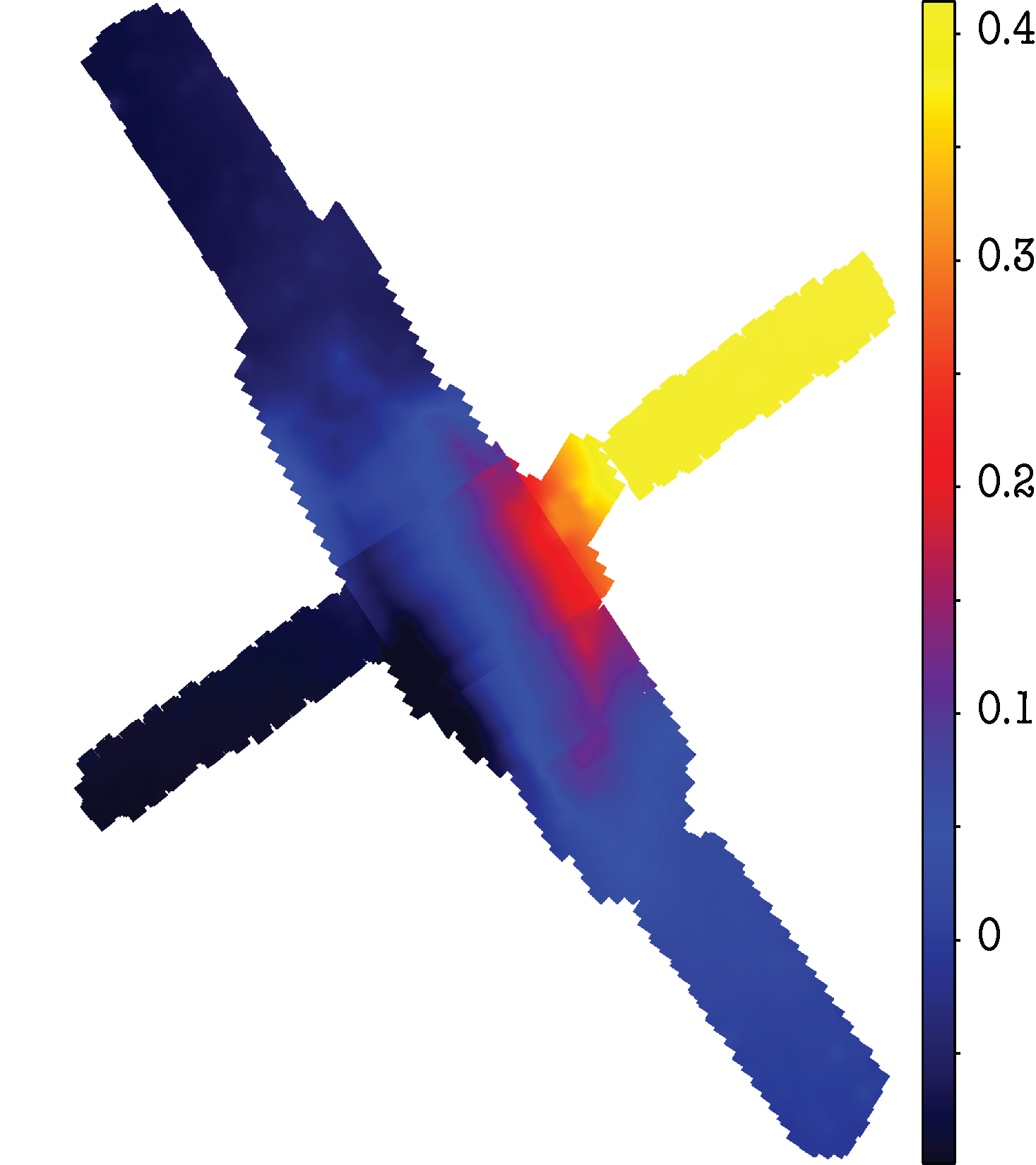}
\caption{Background model at 3.6\,\micron\,.
  North points up and East is towards left.
  The colourbar [MJy~sr$^{-1}$] is constrained and scaled to show background variations.}
\label{fig:bg_colour}
\end{figure}

Using mosaics where galaxies and other sources were masked, background maps were produced as follows. A median value was computed in square regions
of 100$\times$100 pixels that ultimately cover the full mosaics. Fully-masked regions were assigned
an average value based on the median in their neighbouring regions. Then, all pixels were replaced with corresponding
median values in their region. The galaxy, masked by an ellipse, was assumed to follow a first-degree polynomial over
rows and columns. Straight lines were fit to end points along the rows and columns, hence across the elliptical mask
over the galaxy. The background maps were smoothed by taking the average value in a moving box of 199$\times$199 pixels.
Finally, the smoothed maps were subtracted from the calibrated mosaics. Fig.~\ref{fig:bg_colour} shows
the 3.6\,\micron\ background model which is very similar to that for the 4.5\,\micron\ background.
The raw mosaics and modelled background images will be available at \url{http://irsa.ipac.caltech.edu/data/SPITZER/M31IRAC}.

\section{IRAC Catalogue}
\subsection{IRAC catalogue construction}
\label{sec:irac_cat}
The extended IRAC maps contain a wealth of information not only in the pixel-by-pixel light variation but also through
distinct properties of point sources. Point sources in the mosaics were extracted using {\sc{SExtractor}} in dual-image mode, with the 3.6\,\micron\ image used for detection; the 3.6 and 4.5\,\micron\ images
were used for photometry. Point sources were extracted on the mosaics before the background modelling and subtraction
described in section~\ref{sec:bg_sub}; large-scale background subtraction should not affect point-source photometry
since {\sc{SExtractor}} determines the background local to each object.
Input parameters for {\sc{SExtractor}} were set through experimentation and visual inspection,
with the final values given in Table~\ref{tab:settings}.

\begin{table}
\centering
\caption{Parameter settings for {\sc{SExtractor}}}
\label{tab:settings}
\begin{tabular}{lccr}
\hline\hline
Parameter & Value\\
\hline
\texttt{DETECT\char`_MINAREA} [pixel] & 3 \\
\texttt{DETECT\char`_THRESH} & 1.0 \\
\texttt{FILTER} & Y \\
\texttt{FIlTER\char`_NAME} & \texttt{gauss\char`_1.5\char`_3x3.conv} \\
\texttt{DEBLEND\char`_NTHRESH} & 64 \\
\texttt{DEBLEND\char`_MINCONT} & 0.0001 \\ 
\texttt{SEEING\char`_FWHM} [arcsec] & 1.66  \\
\texttt{GAIN} & 0.0 \\ 
\texttt{BACK\char`_SIZE} [pixel] & 64 \\
\texttt{BACK\char`_FILTERSIZE} & 3 \\
\texttt{BACKPHOTO\char`_TYPE} & \texttt{LOCAL} \\
\texttt{BACKPHOTO\char`_THICK} & 24  \\ 
\texttt{WEIGHT\char`_TYPE} & \texttt{MAP\char`_WEIGHT}\\
\hline
\end{tabular}
\end{table}

{\sc{SExtractor}} could not process the full mosaic images due to memory limitations,
so we ran it on sub-sections of the mosaics
(larger sub-sections in the uncrowded outer regions, and smaller sub-sections closer to the disc)
and combined the resulting sub-catalogues.
The coverage images generated during mosaicing were used
as `weight maps' input to the {\sc{SExtractor}} detection procedure, such that a faint object appearing
on a deeper area of the image receives greater weight than one on a shallower area.
The area measured includes all of the area covered by the Cycle 8 data as well as some of the disc region
from the Cycle 1 observations. 

Photometry with {\sc{SExtractor}} was performed in circular apertures of radius 
1\farcs5, 2\farcs0, 2\farcs5, 3\farcs0, and 6\farcs0 as well as in the \texttt{AUTO} and \texttt{ISOCOR} system.
Because {\sc{SExtractor}}'s photometric uncertainties do not account for correlated noise between
mosaiced pixels, we account for this by multiplying the uncertainties by a factor of two, as in \citet{boyer15}.

\subsection{Catalogue description}
The IRAC M31 catalogue is presented in Table~\ref{tab:cat}.\footnote{Also available at \url{http://irsa.ipac.caltech.edu/data/SPITZER/M31IRAC}.}
To avoid noisy regions near the edges, only mosaic regions with coverage$\,\geq\,$2  images per sky position at 
3.6\,\micron\ were used to generate the catalogue. This is a total area of about 15,631\,arcmin$^2$ (4.342\,deg$^2$).
The catalogue contains 426,529 point sources detected at 3.6\,\micron\ of which 423,588 are also detected at 
4.5\,\micron.
Aperture corrections were derived from \citet{ashby09}
and are also given in Table~\ref{tab:apcor}.
The magnitude uncertainties given do not include systematic calibration uncertainty \citep[2~per~cent;][]{reach05}. 
The saturation limits for 30-second frames are 10 and 12\,mJy in 3.6 and
4.5\,\micron\ bands \citep{irac_dhb}, or  $m_{\rm Vega}\,$$=\,$11.1 and 10.4, respectively.
The analysis in the remainder of this paper uses IRAC aperture magnitudes, measured in a
2\arcsec\,radius aperture, large enough to contain a majority ($\sim\,$60~per~cent)
of the light from point sources but not so large as to inflate the photometric noise from background; in addition, all used point sources (including those in auxiliary catalogues) have magnitude uncertainties of less than 0.2.

\begin{table*}
\centering
\caption{IRAC M31 catalogue column definitions.
This table will be available in its entirety in a machine-readable form at \url{http://irsa.ipac.caltech.edu/data/SPITZER/M31IRAC}. A portion is shown here for guidance regarding its form and content.}
\label{tab:cat}
\begin{tabular}{lccccr}
\hline\hline
Column & Parameter         & Description               & Units\\
\hline
   0 & \texttt{ID}           &   M31IRAC JHHMMSS.ss$+$DDMMSS.s & \\
   1 & \texttt{Number}   &       Running object number & \\
   2 & \texttt{KRON\char`_RADIUS}  &   Kron apertures in units of A or B & \\
   3 &  \texttt{A\char`_IMAGE}      &   Profile RMS along major axis             &       [pixel] \\
   4 &\texttt{X\char`_IMAGE}     &    Object position along x                &         [pixel] \\
   5 & \texttt{Y\char`_IMAGE}   &      Object position along y            &             [pixel] \\
   6 & \texttt{X\char`_WORLD}    &     Barycenter position along world x axis    &      [deg] \\
   7 & \texttt{Y\char`_WORLD}   &      Barycenter position along world y axis     &     [deg] \\
   8 & \texttt{ALPHA\char`_J2000}   &  Right ascension of barycenter (J2000)    &       [deg] \\
   9 & \texttt{DELTA\char`_J2000}   &  Declination of barycenter (J2000)      &         [deg] \\
  10 & \texttt{ELLIPTICITY}   &  1 $-$ \texttt{B\char`_IMAGE/A\char`_IMAGE} & \\
  11 & \texttt{FWHM\char`_WORLD}    &  FWHM assuming a Gaussian core   &                [deg] \\
  12 & \texttt{CLASS\char`_STAR}  &    S/G classifier output & \\
  13 & \texttt{FLAGS}      &     Extraction flags & \\
\dots & \texttt{MAG\char`_AUTO\char`_i}        & Kron magnitude                       & Vega mag\\  
\dots & \texttt{MAGERR\char`_AUTO\char`_i}     & Kron magnitude uncertainty           & Vega mag  \\
\dots & \texttt{MAG\char`_APER\char`_i}        & ap mags: 1\farcs5, 2\arcsec, 2\farcs5, 3\arcsec, 6\arcsec\,radii  & Vega mag  	\\    
\dots & \texttt{MAGERR\char`_APER\char`_i}     & aperture magnitude uncertainties     & Vega mag \\
\dots & \texttt{MAG\char`_ISOCOR\char`_i}      & isophotal mag above det threshold    & Vega mag\\
\dots & \texttt{MAGERR\char`_ISOCOR\char`_i}   & isophotal magnitude uncertainty      & Vega mag  \\
\dots & \texttt{FLUX\char`_AUTO\char`_i}        & Kron flux                       & \ujy \\  
\dots & \texttt{FLUXERR\char`_AUTO\char`_i}     & Kron flux uncertainty           & \ujy  \\
\dots & \texttt{FLUX\char`_APER\char`_i}        & ap flux: 1\farcs5, 2\arcsec, 2\farcs5, 3\arcsec, 6\arcsec\,radii  &   \ujy	\\    
\dots & \texttt{FLUXERR\char`_APER\char`_i}     & aperture flux uncertainties     & \ujy \\
\dots & \texttt{FLUX\char`_ISOCOR\char`_i}      & corrected isophotal flux    & \ujy \\
\dots & \texttt{FLUXERR\char`_ISOCOR\char`_i}   & isophotal flux uncertainty      &  \ujy \\
\dots & \dots & \dots \\
  78 & \texttt{P(M31)}     &     Probability for being M31 object       & \\
  79 & \texttt{N\char`_ROBUST} &       Robust count = $P(M31)$$\,\times\,($Completeness$)^{-1}$ & \\
  80 & \texttt{R\char`_PROJ}     &     Projected galactocentric distance     &          [arcmin] \\
  81 & \texttt{R\char`_DEPROJ}    &    Deprojected galactocentric distance   &          [arcmin] \\
\hline
\end{tabular}
\end{table*}

\begin{table}
\centering
\caption{Aperture corrections for M31 IRAC mosaics -- corrections to be added to 
aperture magnitudes to convert them to 
total magnitudes.}
\label{tab:apcor}
\begin{tabular}{lccccccr}
\hline\hline
Band & \multicolumn{5}{c}{Aperture radius}\\
 & $1\farcs5$ & $2\farcs0$ & $2\farcs5$
& $3\farcs0$ & $6\farcs0$\\
\hline
3.6 & $-0.67$ & $-0.38$ &$-0.24$&$-0.17$ & $0$\\ 
4.5 & $-0.69$ & $-0.40$ &$-0.25$&$-0.17$ & $0$\\
\hline
\end{tabular}
\end{table}

\subsection{Completeness}
Catalogue completeness was analysed through the usual `artificial star' method on the 3.6\,\micron\ mosaics.
A point-spread function (PSF) was simulated by making cutout images of about 90 bright
(13$\,\gtrsim\,$[3.6]$\,\gtrsim\,$11),
isolated stars from the mosaic, scaling the images to the same total flux, and co-adding them weighted by star magnitude.
Although the PSF varies over the mosaic due to the spatially-varying depth and position angle,
we verified that the completeness results are not strongly dependent on the exact PSF shape.
Artificial stars were inserted into the mosaic images,  {\sc{SExtractor}} was run on the resulting images in the
same manner as for the real source, and the artificial stars identified in its output based on their known input positions.
About 83,000 artificial sources were inserted at random positions in
the same regions of the major- and minor-axis mosaics used for analysis,
with a power-law ($\alpha\,$=\,0.3) distribution of magnitudes in the range 14$\,<\,$[3.6]$\,<\,$22.
The artificial stars were inserted 5,000 at a time in the uncrowded outer regions of the mosaic,
and 500 at a time in the inner regions, 
following the same sub-region procedure as for the real catalogue.
An artificial star was considered to be detected if its {\sc{SExtractor}}-measured position was within
a distance of 1.5 pixels from the input position. To account for the effects of crowding,
artificial objects were also required to be more than 2\farcs5 from any real catalogue sources,
and have magnitudes within 1.0~mag of the input value, to be considered detected.

\begin{figure}
\centering
\includegraphics[height=84mm]{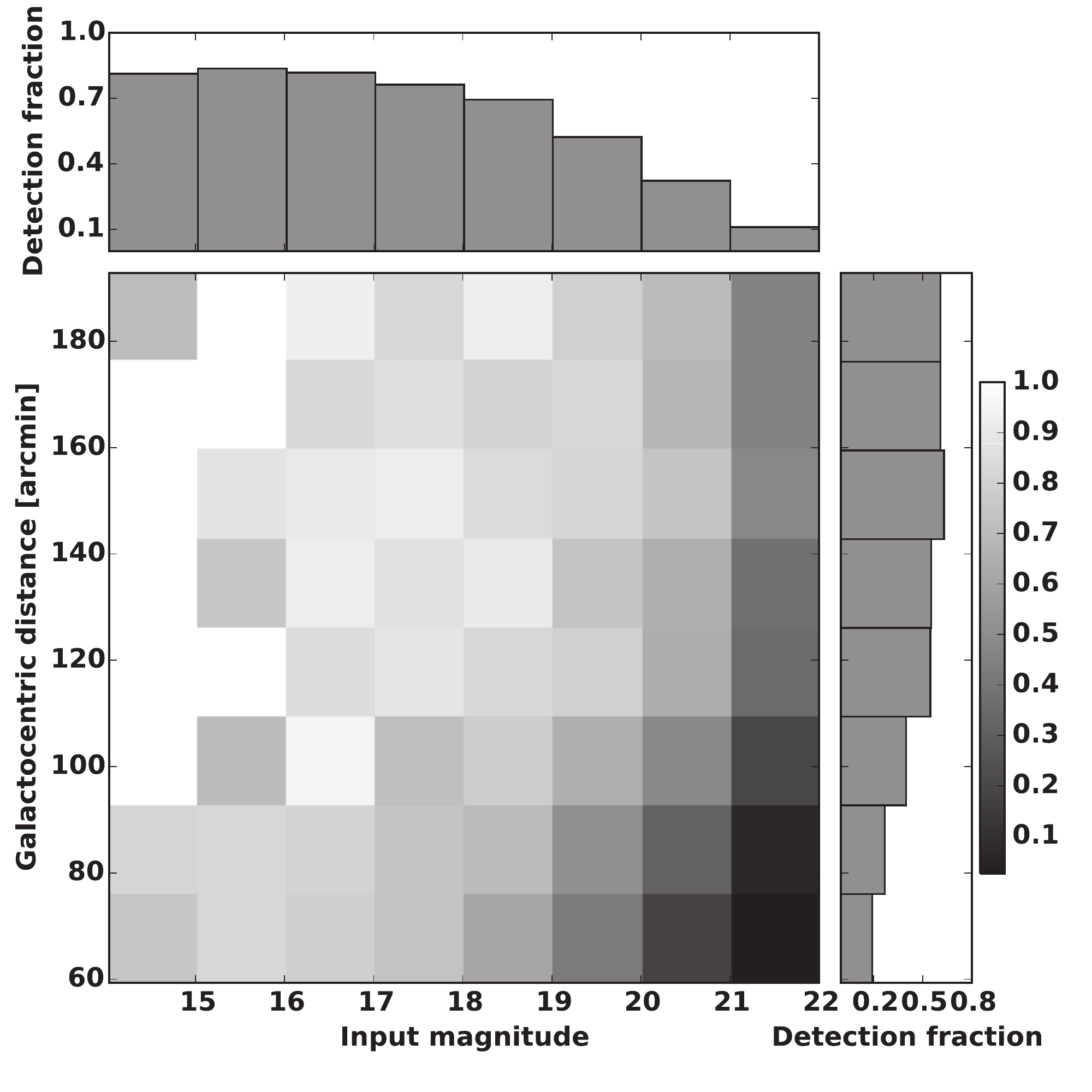}
\caption{Completeness for M31 IRAC catalogue, as derived from artificial star tests on the 3.6\,\micron\ major-axis mosaic.
  Completeness is defined as the fraction of input objects in a given bin of location (projected galactocentric distance) and 3.6\,\micron\ magnitude
  detected within 1.5 pixels and 1.0 magnitude of the input values, and not found within
  2\farcs5 of a real catalogue source.
  Histograms at top and right show the two-dimensional completeness histogram summed along
  the other dimension.
  A similar analysis (not shown) was carried out for the catalogue of sources on the 3.6\,\micron\ minor-axis mosaic.
  } 
  \label{fig:complete}
\end{figure}

Fig.~\ref{fig:complete} shows the completeness estimates derived by
sorting the artificial sources into bins by input magnitude and projected galactocentric distance,
then dividing the number of recovered sources by the number input in each bin.
As expected, the completeness declines as artificial objects become fainter and/or closer to
the galaxy disc. Even for the brightest sources, the catalogue is not 100~per~cent complete, because
a bright artificial star can randomly fall too close to another bright source to be a separate detection.
The 50~per~cent completeness limit is [3.6]\,=\,18.5 at R\,=\,17\arcmin\,and [3.6]\,=\,21.3 at R\,=\,180\arcmin, where radial distances are projected.
Testing {\sc{SExtractor}}'s photometry by comparing input and recovered magnitudes,
we found that the offsets between input and output magnitudes were consistent with zero.
The exception is in the innermost bins of projected galactocentric distance, where output magnitudes were
brighter than input, presumably due to the effects of crowding.
We have not attempted to quantify the reliability of the catalogue (i.e., the fraction
of spurious noise-induced sources) as we believe that crowding-induced incompleteness is likely to be
a much larger effect than noise spikes being mistaken for real sources. 

\subsection{Checks on astrometry and photometry}
\label{sec:astro_phot_precision}
Cross-matching with other catalogues is one way to characterise the photometric and astrometric
precision and accuracy of the IRAC catalogue.
The {\it{WISE}} All-Sky Survey \citep{wise} included the Andromeda region at 3.4\,\micron\ ({\it{W}}1), 4.6\,\micron\ ({\it{W}}2), 12\,\micron\ ({\it{W}}3) and 22\,\micron\ ({\it{W}}4) with angular resolutions of 6\farcs1, 6\farcs4, 6\farcs5, 12\farcs0, and its catalogue of point sources is publicly available in the IRSA\footnote{\url{http://irsa.ipac.caltech.edu/cgi-bin/Gator/nph-scan?mission=irsa&submit=Select&projshort=WISE}}.
{\it{WISE}} detection limits are {\it{W}}1$\,<\,$15.3, {\it{W}}2$\,<\,$14.4, {\it{W}}3$\,<\,$10.1 and {\it{W}}4$\,<\,$6.7~Vega mag at SNR\,=\,5.
We obtained an IRAC/{\it{WISE}} catalogue of point sources by finding best matches within a 2\arcsec\,radius.

With its larger point-spread function, {\it{WISE}} data will be more affected by crowding than IRAC, so we carry out the following comparisons only in the 
outer regions (where R$_{\rm deproj}$\,$>$\,110\arcmin) of the IRAC dataset.
Our matching of IRAC and {\it{WISE}} sources shows that the mean and standard deviation of the separation between matched sources
is $0\farcs1\pm0\farcs4$ in RA and Dec. This is consistent with the expected IRAC positional uncertainties
 \citep{irac_dhb} and indicates that the mosaicing process did not substantially worsen the astrometry.
In uncrowded fields {\it{WISE}} photometry {\it{W}}1 and {\it{W}}2 has been shown to be within 3~per~cent of IRAC 3.6 and 4.5\,\micron\ \citep{Jarrett11}.
We compared aperture photometry for all matched, non-saturated IRAC objects with 11$\,<\,$[3.6\,,\,4.5]$\,<\,$15, 
and found  mean and standard deviation of photometric offsets (IRAC$-${\it{WISE}}) 0$\pm$0.01~mag in both bands. 
Photometry on the IRAC mosaics is therefore inferred to be reliable.

\section{Source Characterization}
\subsection{Catalogue object properties}
We now explore the properties of the individual point sources detected in the M31 mosaics. Three classes of objects
are expected to dominate the population: foreground Milky Way (MW) stars, bright M31 stars, and background galaxies.
\citet{boyer15} give a detailed discussion of the stellar populations that dominate the resolved 3.6 and 4.5\micron\ 
light: both red giants and asymptotic giant branch stars are important. While the spectral energy distribution of 
most stars is Rayleigh-Jeans in the 3.6 and 4.5\micron\ bands and thus their Vega-magnitude colours are [3.6]$-$[4.5]$\,\sim\,$0, stars with circumstellar dust can have colours as red as [3.6]$-$[4.5]$\,\sim\,$1.0.
The brightest M31 objects are expected to be AGB stars with an absolute magnitude of M$^{\rm AGB}_{3.6}=-10.3$ \citep{boyer09}.
This translates into an apparent magnitude of [3.6]$\,\geq\,$14.1,
which we choose as the bright magnitude limit for potential M31 objects through the rest of our analysis.
The tip of the red giant branch in M31 is expected to be about 4 magnitudes fainter than the brightest AGB stars \citep{boyer15},
or [3.6]$\,\approx\,$18. Foreground Milky Way dwarf stars are expected to be dwarfs with  [3.6]$-$[4.5]$\,\sim\,$0 (and see section~\ref{sec:fg_stars}).

\begin{figure}
\centering
\includegraphics[height=78.5mm]{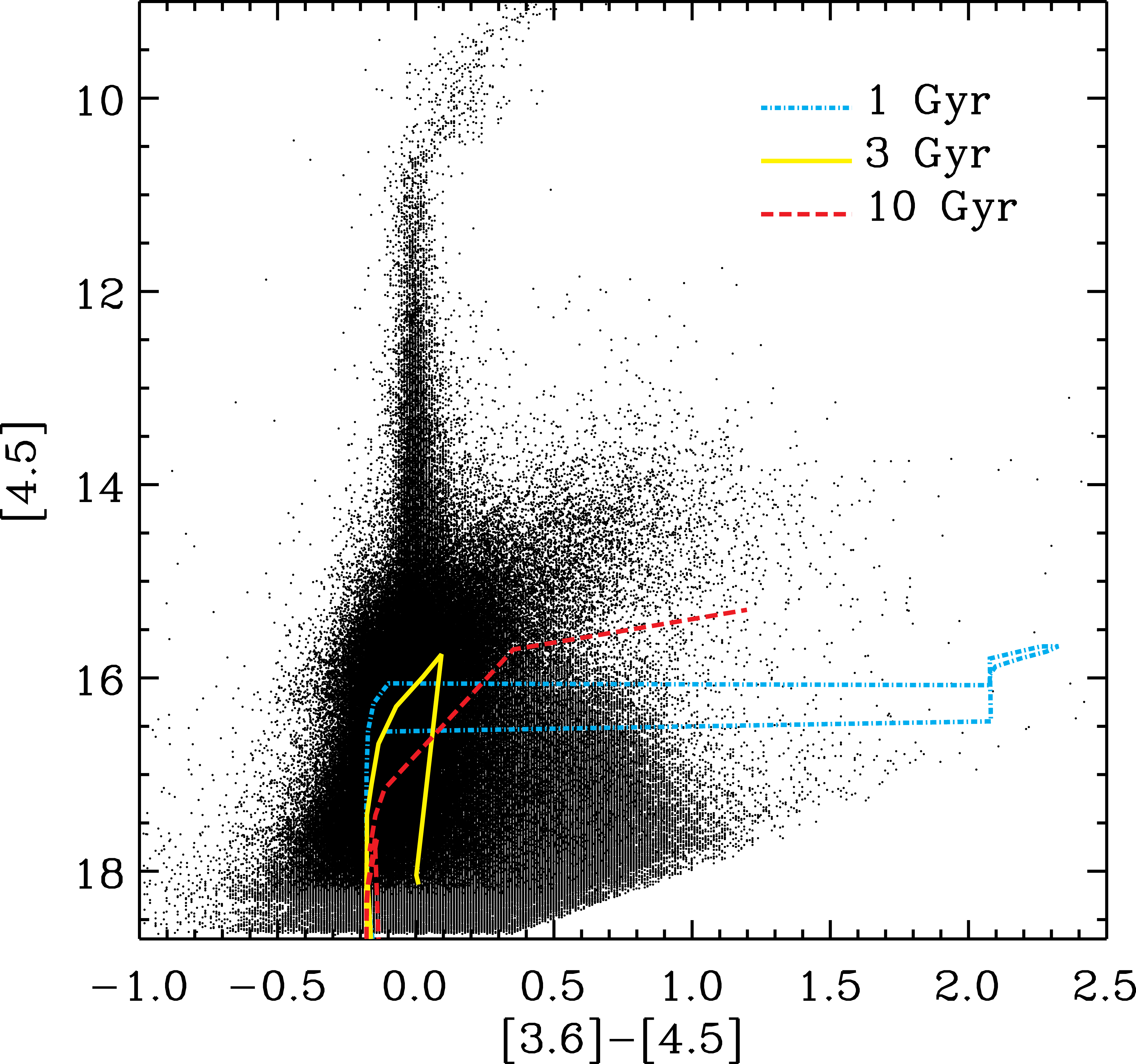}
\caption{The mid-infrared CMD of IRAC sources overlaid with isochrones from \citet{marigo08}.
  A metallically of $Z=0.016$ is assumed for ages of 1, 3, and 10\,Gyr. The red plume of objects at $[3.6]-[4.5]\simeq1$ are likely to be associated with M31.
  The horizontal branch at $[4.5]<10$ is artificial and due to the saturation limits of the IRAC passbands.
  Faint objects beyond $[3.6]-[4.5]>1$ are likely to be red background galaxies. All point sources have magnitude uncertainties of less than 0.2.}
  \label{fig:isoc}
\end{figure}

Background galaxies are expected to have a broad range of colours and magnitudes; background galaxy colours
and luminosity functions can be estimated by comparison with blank-field extragalactic surveys.
The brighter background galaxies can be identified with the {\sc{SExtractor}} \texttt{CLASS\char`_STAR} output
variable \citep{egs}. Selecting sources with [3.6]$\,<\,$15 and \texttt{CLASS\char`_STAR}$\,<\,$0.05, we identified approximately 3,882 potential galaxies in the catalogue.
Further discussion of background galaxies is found in section~\ref{sec:bg_glx};
as these sources are not the focus of this work,
we have not attempted to analyse them in detail (e.g., by measuring shapes or applying extended source photometric corrections).

\begin{figure*}
\centering
\includegraphics[height=79mm]{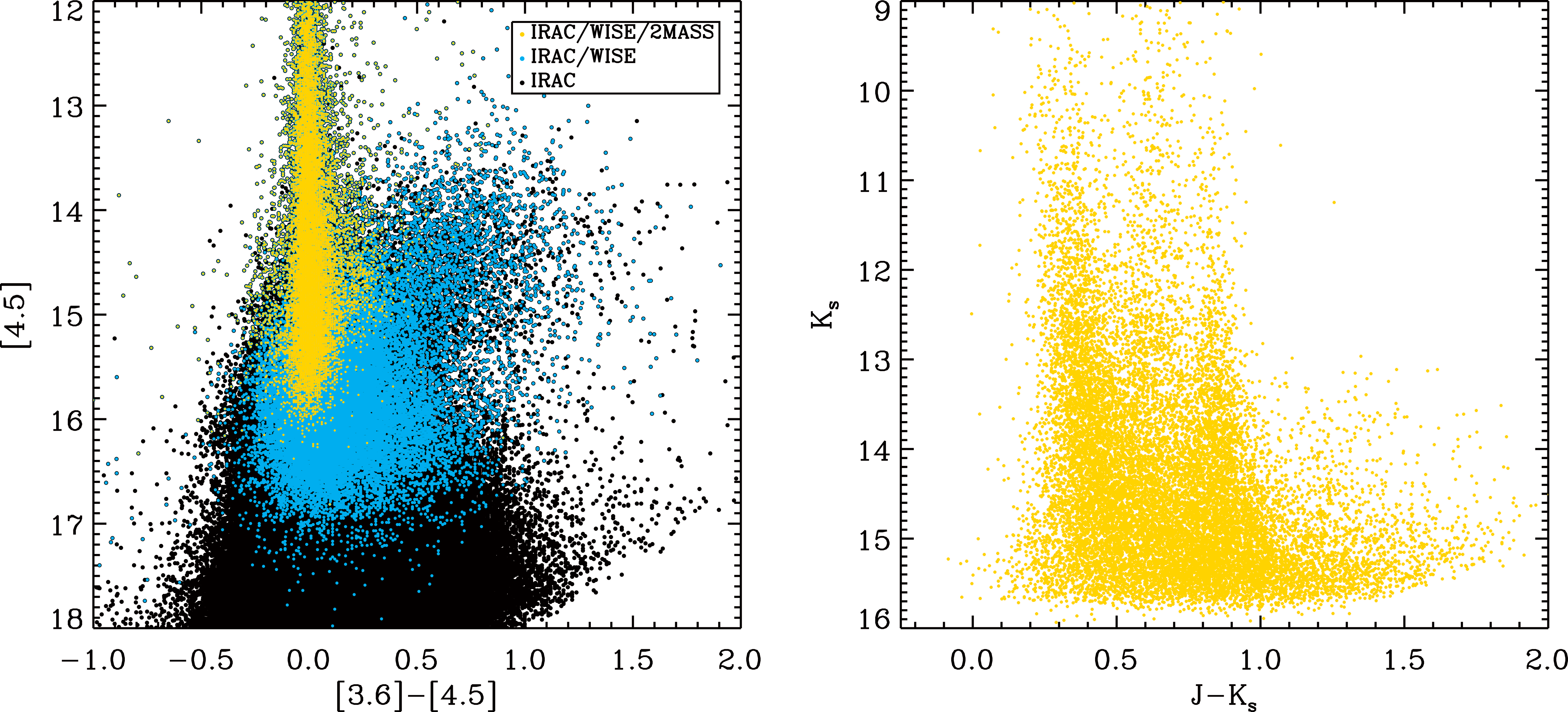}
\caption{Colour-magnitude diagrams of cross-matched IRAC/{\it{WISE}} point sources with a tolerance of 2\arcsec\,. Black points in the background show (un)matched sources in the IRAC catalogue. All point sources have magnitude uncertainties of less than 0.2.
  Left: IRAC-band CMD. Right: Near-infrared (2MASS) CMD of IRAC/{\it{WISE}} sources that also have robust counterparts in the 2MASS catalogue.
  }
\label{fig:irac_wise}
\end{figure*}

Fig.~\ref{fig:isoc} shows the colour-magnitude diagram (CMD) of the IRAC catalogue overlaid with isochrones from \citet{marigo08}\footnote{\url{http://stev.oapd.inaf.it/cgi-bin/cmd}} with the corrections from Case A in \citet{girardi10}, bolometric corrections
from \citet{boloc2} and \citet{boloc1}, and circumstellar dust models from \citet{matg06}.
We use the default values: a metallicity of $Z\,$=\,0.016 coupled with a circumstellar
dust composition of 60~per~cent silicate and 40~per~cent AlOx (amorphous porous Al$_2$O$_3$) are assumed for ages 1, 3, and 10\,Gyr.
The total extinction is assumed to be zero.
Most of the stars and isochrones are concentrated around
the [3.6]$-$[4.5]$\,\sim\,$0 colour; however, a redder branch predicted in the 10\,Gyr isochrone is also
apparent in the data.

\subsection{Multi-wavelength identification}
\label{sec:aux_cat}
The Andromeda neighbourhood has been well-observed in other passbands using both ground- and space-based telescopes, 
providing the opportunity for cross-correlation and identification of sources.  When comparing
the IRAC photometry with other measurements it should be kept in mind that
bright AGB stars often are photometric variables, with amplitudes of a magnitude
or more. If the photometry is not recorded at the same epoch then agreement will
suffer as stars near the faint limit move in and out of detectability. This is likely to be of greatest concern for the reddest AGB stars when making comparison
with visible-light observations, which will be biased against detecting the reddest, most evolved
stars due to line blanketing effects and obscuration by circumstellar discs.
Even for variable stars which are well-detected with IRAC,  the
combination of the instrument's offset fields of view and the observing strategy
means that 3.6 and 4.5\,\micron\ observations of a given object are generally not
simultaneous. This will introduce additional scatter into the colours of variable sources.

Cross-matching between IRAC and the {\it{WISE}}
All-Sky Survey was introduced in section~\ref{sec:astro_phot_precision}.
15~per~cent of IRAC sources have robust (i.e., sources with magnitude uncertainties of less than 0.2) {\it{WISE}} counterparts over our extended observations.
37~per~cent of IRAC/{\it{WISE}} sources also have robust matched counterparts from the Two Micron All-Sky Survey \citep[2MASS,][]{2MASS},
which operated at 1.25\,\micron\ ($J$), 1.65\,\micron\ ($H$), and 2.17\,\micron\ ($K_s$) with an angular resolution of
2\arcsec\,in each of the three bands. 2MASS detection limits are $J$$\,<\,$15.8, $H$$\,<\,$15.1 and $K_s$$\,<\,$14.3~Vega mag at SNR\,=\,10.
Fig.~\ref{fig:irac_wise} shows colour-magnitude diagrams of cross-matched IRAC/{\it{WISE}} point sources.

The {\it{WISE}} dataset is relatively shallow, and comparison with a deeper mid-infrared survey can give insight
into the types of objects present in the M31 catalogue, including contaminating foreground and background sources (see sections~\ref{sec:fg_stars} and \ref{sec:bg_glx}).
Visually inspecting and comparing our M31 CMD to that from M33 observations by \citet{mcquinn07}, we identified the two main populations of M31 objects in mid-infrared: oxygen-rich giant and carbon stars with approximate mean [3.6]$-$[4.5] colours of $-0.1$ and $0.3$, respectively. While the `blue' giants are
more numerous, the redder carbon stars contain a significant number of variable stars -- \citet{mcquinn07} reported 34~per~cent for the red plume of objects in the M33 CMD.

\begin{figure}
\centering
\includegraphics[height=78.5mm]{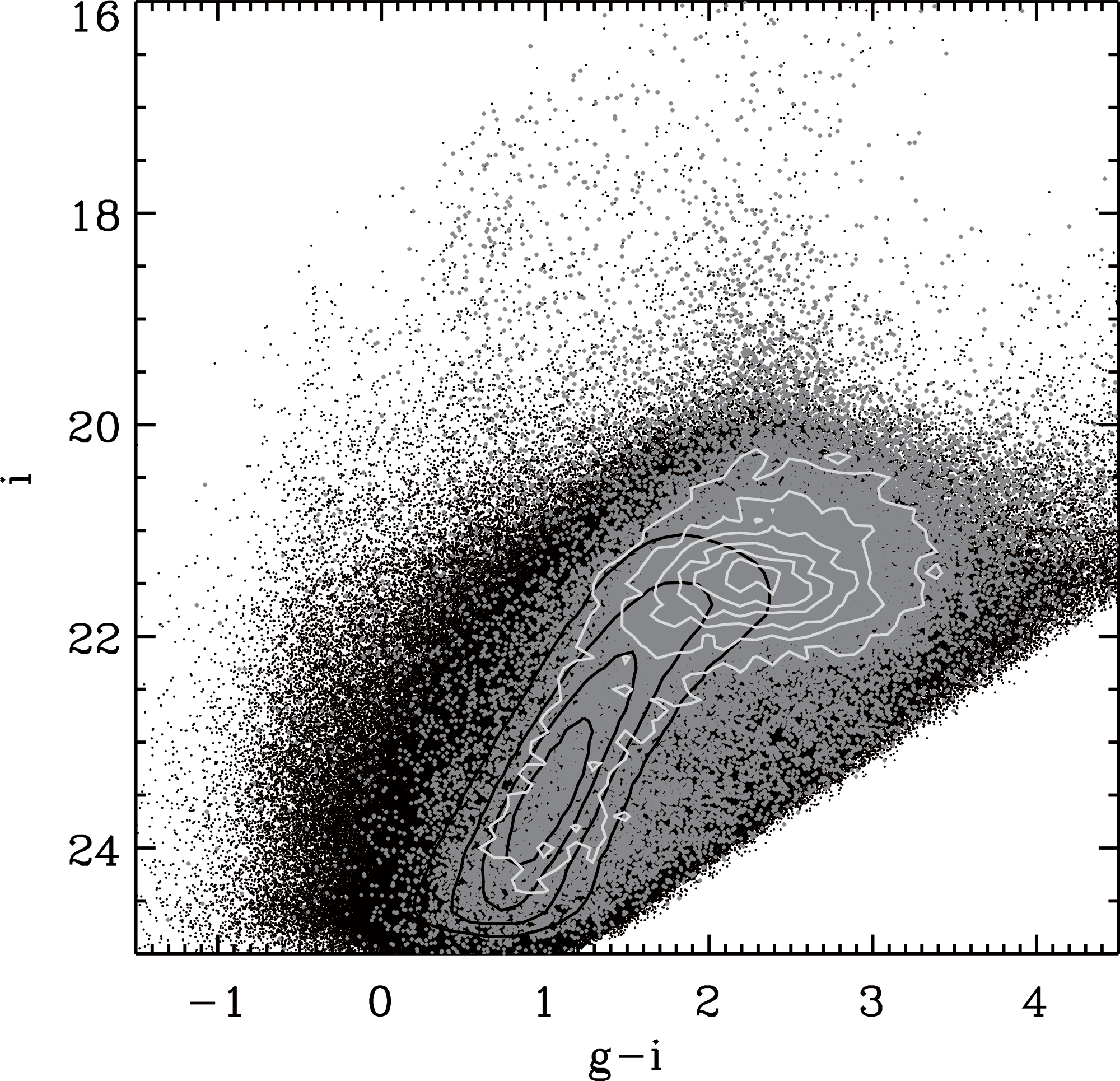}
\caption{CMD of cross-matched point sources with a tolerance of 1\arcsec\,in the visible regime.
  The selected auxiliary (PAndAS, in black) coverage is slightly larger than the full IRAC (in grey) observations.
  The linear contours are to illustrate number density variations using a CMD bin size of 0.1$\times$0.1.}
\label{fig:pand}
\end{figure}

The Pan-Andromeda Archaeological Survey \citep[PAndAS,][]{pandas} covered over $\sim\,$400\,deg$^2$
of the Andromeda neighbourhood using CFHT/MegaCam in $g$~(4,140 -- 5,600\,\AA) and $i$~(7,020 -- 8,530\,\AA) passbands with
mean resolutions of $0\farcs67$ and $0\farcs60$, respectively. The PAndAS catalogue has median $5\sigma$ detection limits of
26.0 and 24.8~Vega mag in the $g$ and $i$ bands, respectively.
Fig.~\ref{fig:pand} shows the CMD of PAndAS sources that have (not) been matched to IRAC sources within a 1\arcsec\,radius;
IRAC sources are clearly centred around two lobes in the red regime, indicating very mid-infrared-bright objects
(e.g., giant stars and background galaxies) that extend their SED to the visible regime.

\subsection{Estimating contamination: Milky Way stars}
\label{sec:fg_stars}
Tracing the light of M31 requires a careful treatment of all possible sources of contamination due to
the foreground Milky Way stars as well as background galaxies. While all-sky surveys can be helpful in
estimating the contribution from foreground objects at a given Galactic latitude,
there still remains the possibility of cross-contamination from the M31 halo.
This unwanted effect can be mitigated by choosing foreground samples well outside the halo.
Using the {\it{WISE}} All-Sky Survey and {\sc{trilegal}} \citep[v1.6,][]{trilegal} stellar population models we obtained the probability \citep{bagheri} that a point source would belong to M31, by comparing to the number density of foreground stars:
\begin{equation}\label{eq:prob}
P(M31) = \frac{N_{M31} - N_{MW}}{N_{M31}}
\end{equation}
where $N$ is the number of stars in a CMD bin. Highly contaminated bins with $N_{MW}$$\,\geq\,$$N_{M31}$ were assigned a probability value of zero; thus, $0\leq P(M31) \leq1$.
Using the {\it{WISE}} catalogue we selected two 1\degr$\times\,$1\degr\ foreground regions at ($23^{\rm h}29^{\rm m}20^{\rm s}, +37\degr46\arcmin35\arcsec$)
and ($03^{\rm h}08^{\rm m}03^{\rm s}, +32\degr26\arcmin10\arcsec$), which are well outside our extended IRAC coverage but
within $-22\degr\leq b\leq -21\degr$. We linearly scaled their number densities according to the total area covered by
the IRAC observations. Assuming a homogeneous distribution of Milky Way stars along the same Galactic latitude, our foreground fields are representative samples of stellar populations in the M31 neighbourhood. Using these foreground regions and the matched IRAC/{\it{WISE}} catalogue, we used equation~(\ref{eq:prob})
to compute probability values in CMD bins with [4.5]$\,\le\,$16. This
approach is independent of the instruments' detection efficiencies as
it only involves matched point sources between the two catalogues. The
{\it{WISE}} catalogue does not provide a statistically robust sample of objects for bins fainter than [4.5]$\,\sim\,$16. 

To model the fainter foreground stars, we used stellar population models to
estimate the distribution of MW stars with [4.5]$\,>\,$16 at the
position of M31. Using the same extended IRAC coverage we generated a
{\sc{trilegal}} run with default values.
Next, model magnitudes were smoothed by adding a Gaussian noise from
the IRAC uncertainties -- we chose a magnitude bin width of 0.2 and
computed the mean of uncertainties in each bin. 
Observed colour values are more uncertain than magnitude values because
they combine two measurements; for the colour width of the bins we chose a value of 0.8.
While this is fairly large compared to the CMD width, it ensures
that each bin includes a statistically robust sample of stars and
suppresses the scatter due to variable stars, while still 
mapping visually separable features in the
colour-magnitude space \citep[see, e.g., figure 15 of][]{mcquinn07}.
Using the modified {\sc{trilegal}} catalogue
equation~(\ref{eq:prob}) was used to estimate $P(M31)$ for bins with
[4.5]$\,>\,$16. In this analysis, the faintest objects with magnitude
uncertainties of less than 0.2 have [4.5]$\,\approx\,$18.7.

\subsection{Estimating contamination: background galaxies}
\label{sec:bg_glx}
Another source of contamination is the uniform distribution of background galaxies in mid-infrared. 
Our foreground estimation technique using the {\it{WISE}} catalogue accounts for the contribution from background galaxies with [4.5]$\,\le\,$16. In addition, we selected \emph{stellar} objects by removing sources with [3.6]$\,<\,$15 and \texttt{CLASS\char`_STAR}$\,<\,$0.05 from the IRAC catalogue. The selection criteria for \emph{stellar} objects combined with colour-magnitude uncertainties produced bins with probability values of greater than 1, which were subsequently normalised to 1.

Nevertheless, these criteria fail to identify galaxies fainter than [3.6]$\,\sim\,$16.5 as they are primarily undetected by {\it{WISE}} and also exhibit the same IRAC magnitudes (within the uncertainty range) in all apertures.
For bins with [4.5]$\,>\,$16, we made use of the Spitzer Deep, Wide-Field Survey \citep[SDWFS,][]{ashby09} to statistically measure the abundance of background galaxies in the colour-magnitude bins. Using the IRAC instrument and observing $\sim\,$10.5~deg$^2$ of a relatively foreground-free region at ($l=57\fdg5, b=67\fdg5$), the SDWFS has sufficient depth and coverage to provide a statistically robust sample of background galaxies. Removing the contribution due to any foreground stars, we generated a {\sc{trilegal}} model (with default values) over a 10~deg$^2$ field, at the SDWFS sky position.
Number densities were scaled by the SDWFS coverage and a Gaussian noise was applied to the magnitudes based on the SDWFS magnitude uncertainties.
Using the same bin size and replacing $N_{MW}$ by background galaxy number counts\footnote{In SDWFS colour-magnitude bins, the background galaxy count is given by subtracting the foreground ({\sc{trilegal}}) from the total count.}, equation~(\ref{eq:prob}) was applied to find the probability that a source would belong to MW \emph{or} M31. Multiplying the foreground (based on {\sc{trilegal}} simulations of M31 field) by the background (based on foreground-free SDWFS) probabilities, we combined the statistics for sources with [4.5]$\,>\,$16.

\begin{figure}
\centering
\includegraphics[height=74mm]{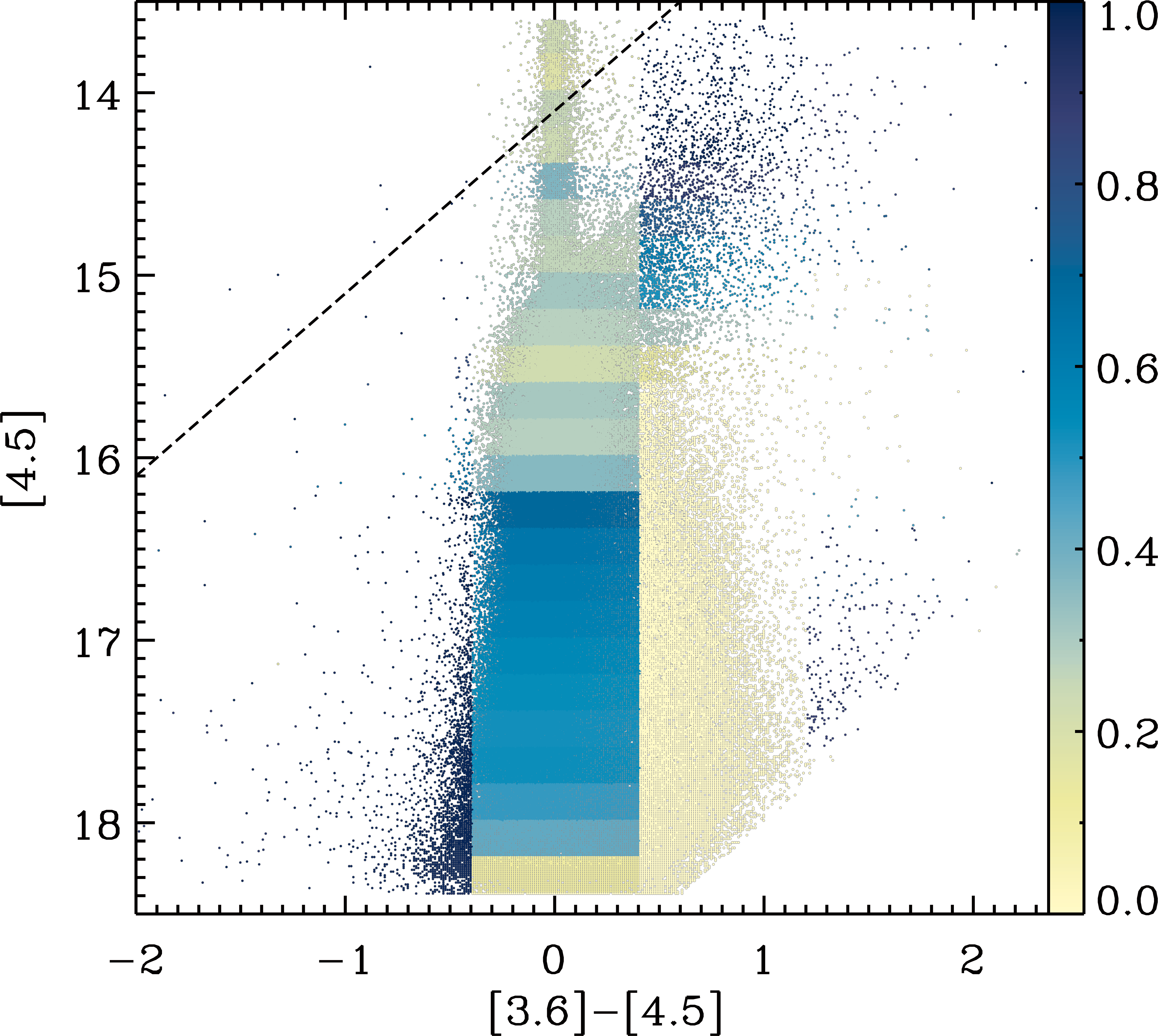}
\caption{The mid-infrared CMD of IRAC sources. The colour-bar represents the probability for a source to belong to M31.
  The dashed line shows the AGB cut, defined by the brightest AGB stars at the distance of M31.
  Objects with [3.6]$\,<\,$14.1 (i.e., above AGB cut) are assumed to be foreground Milky Way stars.
  Highly probable M31 carbon stars coalesce into a clump with [4.5]$\,\gtrsim\,$15 and [3.6]$-$[4.5]$\,\gtrsim\,$0.4. The fainter area under this region is severely contaminated by star-forming background galaxies. The central branch at [3.6]$-$[4.5]$\,\sim\,$0 contains a blend of all sources, including a large population of Milky Way dwarf stars.
  }
\label{fig:cmd_prob}
\end{figure}

The result is shown in Fig.~\ref{fig:cmd_prob}. Attempting to select distinct populations of objects, separated by different colours in multiple datasets, resulted in some sharp discontinuities along the colour axis (i.e., constant [4.5]) -- this is in contrast to the relatively smooth probability variation along the densely-packed magnitude axis. 
As discussed in
section~\ref{sec:fg_stars}, using smaller colour bins would have
introduced other problems, such as variability-induced scatter.
Because of the complexity and uncertainty associated with the
inclusion of multiple auxiliary catalogues, we refrained from
propagating uncertainties in our pipeline; thus, we did not attempt to
smooth out the colour jumps in the CMD. 
 We remind the reader that this is a probabilistic colour-magnitude
diagram based on photometry; we have not attempted to use radial velocities or other cross-identifications to assess M31 membership probabilities.

\subsection{Contamination-corrected catalogue}
After removing bright foreground stars and resolved background galaxies, our M31 catalogue contains
279,509 sources with magnitude depths of [3.6]$\,=\,$19.0$\,\pm\,$0.2 and [4.5]$\,=\,$18.7$\,\pm\,$0.2.
We assigned $P(M31)$ values to each object based on its location in the M31 CMD.
These objects cover a wide range of projected galactocentric radii between $5\farcm4$\ ($22\farcm9$ deprojected) and $200\farcm6$\ ($637\farcm4$ deprojected). We found an average probability value of $P_{\rm avg}(M31)$$\,=\,$0.4$\,\pm\,$0.2, where the uncertainty is empirically  based on the variation of probability values in colour-magnitude space. Cross-matching with other catalogues, we found that the potential M31 sources were robustly matched with 35,437 and 51,574 {\it{WISE}} and PAndAS catalogue sources, respectively. Future work will analyse this rich dataset in detail. In section~\ref{sec:star_cts}, we use the contamination-corrected catalogue to measure number density variations in the disc and halo of M31.

\section{Surface Brightness Profile}
A typical overall approach to determining galaxy surface brightness profiles is to measure integrated light closer to the centre
and combine that with star counts in the outer part. Our large mosaics allow us to perform both measurements for M31 on the same dataset.
The robustness of both background subtraction and catalogue construction can be tested by comparing the two methods
in the radial region where they overlap.

\subsection{Integrated light}
We traced the integrated light by computing the median value in width-varying elliptical bins while keeping the inclination and position angle constant. We used equation~(\ref{eq:wedge}) \citep{courteau11} to determine the exponentially increasing width ($w$) of the elliptical bins:
\begin{equation}
\label{eq:wedge}
w = p\,(r^{n} - 1)
\end{equation}
where $n$ is the bin number starting at 0, $p\,$=\,5\farcm25 and $r\,$=\,1.01 are fine-tuning constants -- these values produce statistically robust samples for tracing the light over size-varying galactic structures (e.g., the 10\,kpc ring). Measuring the flux from M31, we masked out the satellite galaxy NGC\,205 with a square on the minor-axis mosaic.

In each elliptical bin, we obtained a distribution of median values by bootstrapping 100 times over all unmasked pixels.
The bootstrapping was done by sampling a bin's pixels (with replacement) and finding the median for each resampled bin.
A final median value and its uncertainty
were given by the mean and standard deviation of the corresponding distribution for each bin.
To best mitigate possible contamination from bright point sources,
pixels above $+3 \sigma$ level were clipped, and the median was recomputed along with its associated uncertainty.
Furthermore, we computed the uncertainty associated with extreme cases of the background model from section~\ref{sec:bg_sub}. In each bin, we summed the two (i.e., the final standard deviation from bootstrapping and the absolute difference from extreme cases of the background) uncertainties in quadrature. We did \emph{not} use the absolute uncertainty from section~\ref{sec:bg_unc}. The current implementation robustly accounts for much larger uncertainties due to the background and the pixel counts.

For bins larger than 5\arcmin\,in diagonal, median flux values were multiplied by the extended-to-point source
calibration ratios (with a 10~per~cent uncertainty) given in the IRAC Instrument Handbook. Surface brightness in [MJy~sr$^{-1}$] was converted to [mag arcsec$^{-2}$] by the following transformation: $\mu = \mu_{\circ} - 2.5 \log{(I)}$,
where $\mu$ and $I$ are in [mag~arcsec$^{-2}$] and [MJy~sr$^{-1}$], respectively;
$\mu_{\circ}=17.30$ and $16.81$\,[mag~arcsec$^{-2}$] are in the Vega photometric system at 3.6 and 4.5\,\micron\ passbands, respectively.

As a result of uncertainties in the background, bins in 86\arcmin$\,<\,$R$\,<\,$90\arcmin\,produced unreliable measurements; bins beyond R$\,\sim\,$90\arcmin\,contained negative median-flux values, resulting in undefined logarithmic values for the magnitude. We used the distribution of point sources as a light tracer in the outer radii.

\subsection{Star counts}
\label{sec:star_cts}

When the integrated-light profile approaches the background level, surface brightness measurements become uncertain.
In this case, star counts can be used as a tracer of the galaxy light. The inclined distribution of M31 giants inside R\,=\,55\,kpc \citep[see fig.~1 in][]{tanaka10} motivates us to explore number counts in the deprojected bins. To this end, we counted the number of \emph{stellar} objects (see section~\ref{sec:bg_glx}) within the width-varying elliptical annuli from R\,=\,90\arcmin\,(20.6\,kpc) out to R\,=\,200\arcmin\,(45.7\,kpc) -- the distribution of sources outside this range is uncertain due to crowding (in the inner regions) and sparsity (in the outer regions). The elliptical bin width ($w$) was computed by equation~(\ref{eq:wedge}) with $p$\,=\,3\farcm1 and $r$\,=\,1.1; these values provide large counts with small uncertainties for tracing the faint light.
In this work, we only used objects with [3.6]$-$[4.5]$\,>\,$$-0.75$; bluer objects are not physically reliable. The probability values of sections~\ref{sec:fg_stars} and \ref{sec:bg_glx} were included in this analysis,
suppressing the contamination due to the foreground MW stars and background galaxies. Giving a larger weight to \emph{potential} M31 objects, we only considered sources with $P(M31)$$\,\geq\,$0.5; all other sources with $P(M31)$$\,<\,$0.5 were ignored while computing the total counts. In addition, we corrected the counts for incompleteness; the two-dimensional (in magnitude and projected galactocentric distance; see Fig.~\ref{fig:complete}) completeness results were applied, with the same bin parameters as in section~\ref{sec:irac_cat}, to sources detected on the minor- and major-axis mosaics. 

\begin{figure*}
\centering
\includegraphics[height=98mm]{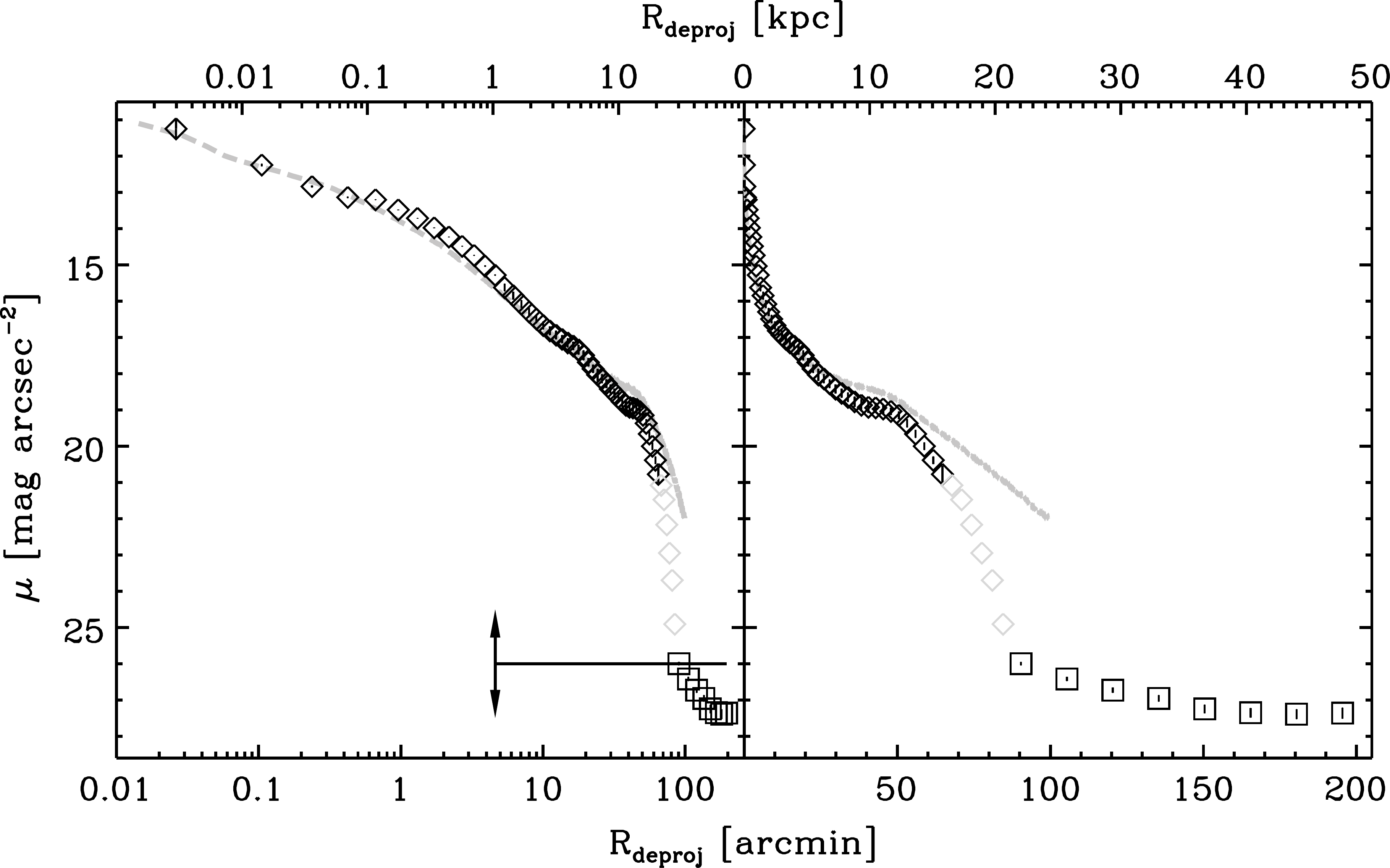}
\caption{M31 surface brightness profiles based on a combination of surface photometry at 3.6\,\micron\ (diamonds) and the radial distribution of stars (squares) over the entire IRAC observations. The grey dashed line shows an azimuthally averaged profile by \citet{courteau11}.
Panels contain identical measurements; a logarithmic axis is used for probing R$\,<\,$20\arcmin\,in the left panel.
The horizontal bar marks $\mu$\,=\,26, which is an arbitrary value (thus shown by arrows) that we chose for the first bin computed by star counts; the following bins were shifted by the same offset.
The break-point between the integrated photometry and star counts occurs at a radial distance of $\sim\,$88\arcmin\,(20\,kpc). Grey diamonds (86\arcmin$\,<\,$R$\,<\,$90\arcmin) are significantly contaminated by the background and have no reliable uncertainties.
}
\label{fig:sfp}
\end{figure*}

In each bin, number densities were converted to surface brightness [mag arcsec$^{-2}$] by the following relation:  $\mu = \mu_{\circ} - 2.5 \log{(N A^{-1})}$, where $N$ is the sum of robust number counts, given by $P(M31)$$\times$(Completeness)$^{-1}$, and $A$ is the enclosed area in [arcsec$^{2}$].
Since number counts do not have an absolute calibration value, we adjusted $\mu_{\circ}$ to fix the first bin at $\mu$\,=\,26.
The Poisson noise model was used for estimating the uncertainties associated with random counts. Including the contribution due to the ignored objects, the absolute difference for extreme cases was added to the Poisson noise in quadrature.
We measured surface brightness profiles at both 3.6 and 4.5\,\micron\ and found the two profiles to be quite similar as also noted by \citet{courteau11}.

Table~\ref{tab:sbp} gives the combined surface brightness profile from
both integrated photometry and star counts.
Fig.~\ref{fig:sfp} shows the star count profile that has been shifted to illustrate the continuation of the integrated light.
Shifting the star count profile to match the integrated light profile results in
a change in profile slope at the matching point. We suspect that this
reflects the difficulty in determining the appropriate
background levels for the two profiles rather than a true physical
change in profile slope. The availability of point-source photometry with higher
spatial resolution (and thus less affected by crowding) in the overlap region would help to overcome this
difficulty. This should be possible with future facilities such as the
{\it{James Webb Space Telescope}}.

\begin{table}
\centering
\caption{IRAC M31 surface brightness profile. This table serves as a guide for the full table, available in its entirety in a machine-readable form in the online journal. () enclose values with no reliable uncertainties.}
\label{tab:sbp}
\begin{tabular}{lcccr}
\hline\hline
R$_{\rm deproj}$ & median surface brightness & source\\
(arcmin) & (3.6\,\micron\ mag~arcsec$^{-2}$) &\\
\hline
0.02625 & $11.245\pm0.260$ & integrated\\
0.10526 & $12.244\pm0.037$ & integrated\\
0.23757 & $12.838\pm0.028$ & integrated\\
\dots & \dots & \dots\\
64.7646 & $20.775\pm0.234$ & integrated\\
67.8535 & $(21.079\pm0.000)$ & integrated\\
\dots & \dots & \dots\\
84.5656 & $(24.909\pm0.000)$ & integrated\\
90.4113 & $26.000\pm0.048$ & star count\\
105.411 & $26.423\pm0.058$ & star count\\
\dots & \dots & \dots\\
195.411 & $27.362\pm0.151$ & star count\\
\hline
\end{tabular}
\end{table}

\section{Discussion and Summary}
We have presented the first results from an extended survey of M31 with {\it{Spitzer}}-IRAC
at 3.6 and 4.5\,\micron. We have produced background-corrected mosaics covering total lengths
of 4\fdg4 and 6\fdg6 along the minor and major axes, respectively. A  3.6\,\micron-selected catalogue of 
point sources is presented for 426,529 objects, many of which are detected for the first time. 
Using auxiliary catalogues we have carried out a statistical analysis in colour-magnitude space
to discriminate M31 objects from foreground Milky Way stars and background galaxies.
These mosaics and catalogue contain a wealth of information for Galactic as well as extragalactic studies
and they are available for other investigators to use.

One of the original goals of this work was to produce surface brightness profiles by combining the integrated light 
from background-corrected maps with statistically-treated star counts. 
Despite the extended coverage of the IRAC mosaics, it is challenging to trace the integrated light for R$\,>\,$86\arcmin; 
the background noise approaches the signal, resulting in very large uncertainty values for the measured flux.
In the case of star counts, we face a similar effect due to the foreground and background contaminants -- Milky Way dwarfs and star-forming background galaxies are mixed with potential M31 objects in the mid-infrared CMD. As a result, bins beyond R$\,\sim\,$200\arcmin\,exhibit no robust fall-off or feature (within the uncertainty range). In the inner regions (R$\,<\,$90\arcmin), star counts suffer from significant crowding and are thus unreliable due to incompleteness.
Because of the difference in sources of contaminants and hence the
difference in methods for overcoming these effects in the two profiles, background levels are prone to inconsistency. Therefore, we refrained from extrapolating and/or normalising profiles in the overlapping region at R$\,\sim\,$88\arcmin.
We conclude that modelling the combined bulge, disc, and halo of the galaxy over a wide range of radii will require combining
the IRAC M31 surface brightness profile with other multiwavelength data. Our integrated light profile is \emph{not} based on any isophotal fitting 
algorithm, and should thus be reproducible and easily comparable to future studies. 

In addition to further work on the galaxy structure, a number of future studies are possible with the extended M31 dataset.
Examples include stellar population studies via comparison with visible-light datasets such as PAndAS \citep{pandas} and PHAT \citep{dalcanton12},
searches for dusty extreme-asymptotic-giant-branch stars as in \citet{boyer15}, comparing a larger sample of M31 globular clusters' mid-infrared colours
to population synthesis models as in \citet{barmby12}, and determining the mid-infrared properties of a uniquely-selected population of
quasars in the M31-background \citep{huo15}. Data-mining of the catalogue for unusual objects which could be followed-up with 
higher-resolution infrared imaging or spectroscopy is another possible direction, as is exploration of stellar variability. The Andromeda galaxy provides us with a 
diversity of physical conditions and star formation histories to explore, with mid-infrared imaging comprising an important spectral region
to fully understand our neighbouring galaxy's story.

\section*{Acknowledgements}
MRR thanks Sahar Rahmani for technical support. We thank Steven~Willner for providing fruitful comments on the initial manuscript; St\'{e}phane Courteau for providing surface brightness profiles from \citet{courteau11}; Alan~W.~McConnachie and the entire PAndAS collaboration for providing access to the PAndAS catalogue. This work is based on observations made with the \emph{Spitzer Space Telescope} and has made use of the NASA/IPAC Infrared Science Archive, both operated by the Jet Propulsion Laboratory, California Institute of Technology, under contract with the National Aeronautics and Space Administration. Support for this work was provided by NASA and by a Discovery Grant from the Natural Science and Engineering Research Council, Canada. This research made use of NASA's Astrophysics Data System; {\sc{montage}}, funded by the National Science Foundation under Grant Number ACI-1440620, and was previously funded by the National Aeronautics and Space Administration's Earth Science Technology Office, Computation Technologies Project, under Cooperative Agreement Number NCC5-626 between NASA and the California Institute of Technology;
{\sc{topcat}}, an interactive graphical viewer and editor for tabular data \citep{taylor05}; the facilities of the Canadian Astronomy Data Centre operated by the National Research Council of Canada with the support of the Canadian Space Agency; and Digitized Sky Survey, retrieved via Aladin \citep{aladin}.
The Digitized Sky Surveys were produced at the Space Telescope Science Institute under U.S. Government grant NAG W-2166. The images of these surveys are based on photographic data obtained using the Oschin Schmidt Telescope on Palomar Mountain and the UK Schmidt Telescope. The plates were processed into the present compressed digital form with the permission of these institutions.\\\\ {\it{Facilities: {Spitzer~Space~Telescope}}}~(IRAC)

\bibliographystyle{mnras}
\bibliography{m31_irac_outer}

\bsp
\label{lastpage}
\end{document}